\documentclass{ws-mpla}

\usepackage{graphics}

\begin{document}

\markboth{Ben-Wei Zhang and Ivan Vitev}
{Direct photon production in d+A and A+A collisions at RHIC }

\catchline{}{}{}{}{}

\title{Direct photon production in d+A and A+A collisions at RHIC}

\author{\footnotesize Ben-Wei Zhang$^{a,b}$ and Ivan Vitev$^a$}

\address{$^a$Los Alamos National Laboratory, Theoretical Division,
Mail Stop B283, Los Alamos, NM 87545, USA }

\address{$^b$Institute of Particle
Physics, Hua-Zhong Normal University, Wuhan 430079, China}

\maketitle


\begin{abstract}
Direct photon production in minimum bias d+Cu and d+Au
and central Cu+Cu and Au+Au collisions at center of mass energies
$\sqrt{s}=62.4$~GeV and 200GeV at RHIC is systematically 
investigated.
We study the jet quenching effect, the medium-induced photon bremsstrahlung 
and jet-photon conversion in the hot QGP. We account for known cold 
nuclear matter effects, such as the isospin effect, the Cronin effect, shadowing  
and cold nuclear matter energy loss. It is shown that at high $p_T$ the
nuclear modification factor for direct photons $R_{AA}^{\gamma} (p_T) < 1$ is
dominated by cold nuclear matter effects and there is no evidence for 
large cross section amplification due to medium-induced photon bremsstrahlung
and jet-photon conversion in the medium. Comparison of numerical
simulations to experimental data also rules out large Cronin
enhancement and incoherent photon emission in the QGP but the 
error bars in the current experimental data cannot provide further
constraints on the magnitudes of other nuclear matter effects. 

\end{abstract}

\ccode{PACS Nos.: {12.38.Mh}{}   \and
      {24.85.+p}{} \and
      {12.38.Bx}{}}

\section{Introduction}
\label{intro} 
In relativistic heavy-ion collisions a large amount of energy 
is deposited in the interaction region over a very short time
interval $\tau \sim 2R_A/\gamma$.
An excited state of matter, the quark-gluon plasma 
(QGP), is expected to be formed and its properties are subject to 
intense theoretical and experimental investigation.
From SPS to RHIC and to the LHC, with increasing $\sqrt{s}$, hard 
probes~\cite{Gyulassy:2003mc,VWZ} have become increasingly more
important as tomographic tools in the study of the QGP. Their 
perturbative QCD (pQCD) calculation is based
on the factorization theorem, which separates the hard partonic part
from the soft, non-perturbative part. For a physics observable $d
F/dyd^2{\bf p}_T$ in p+p collisions we have:
\begin{eqnarray}
\frac{ d F}{ dy d^2{\bf p}_T  } & \propto &
\phi_{a/A}({\xi}_A,\mu_f)
\phi_{b/B}({\xi}_B,\mu_f)\nonumber \\
& &\otimes \frac{ d\hat{F} ^{a+b\rightarrow c + X} }
{ dy d^2{\bf p}_T }\otimes
  D_{h/c}(z,\mu_f) \, .
\label{eq:factorize}
\end{eqnarray}
Here, $ d\hat{F}^{a+b\rightarrow c + X}/dyd^2{\bf p}_T $ represents the
partonic cross section, while $\phi_{a/A}({\xi}_A,\mu_f)$ 
and $D_{h/c}(z,\mu_f)$ are the non -perturbative parton 
distribution functions (PDF) and parton fragmentation functions (FF),
respectively.  In nucleus-nucleus (A+A) collisions, the result of 
final-state interactions can  often be represented as {\em effective}
\footnote{The convenience of a  particular mathematical 
representation of nuclear matter effects on the observable cross section 
should not be confused with genuine universal  modification (which 
may also be present) of the parton distribution and fragmentation functions.} 
modifications of the FFs.  Similarly, cold nuclear matter (CNM) effects
resulting from the initial-state interations in cold nuclei  
can be absorbed in  {\em effective} modifications of the PDFs.
The latter will also be present in A+A collisions and a
robust calculation of QGP signatures needs reliable understanding 
of  initial-state interactions.

Since the photon couples to the partons in the collision
region only through electromagnetic interactions ($\alpha_{em} \ll
\alpha_s$) its mean free path is very large. Hard 
photons leave the medium without rescattering and their production
can provide valuable information to help constrain  initial-state
CNM effects. The study of $\gamma$ production in A+A has so far
been centered on its QGP-specific components, in particular 
medium-induced photon emission in the QGP and jet-photon
conversion~\cite{Srivastava:2008es}.  The predicted consequential
large enhancement of the direct photon production cross section 
 in central heavy-ion collisions appears to be incompatible 
with recent data~\cite{takao}. Our paper~\cite{Vitev:2008vk} 
addresses the need for a systematic study of direct photon
production in both p+A and A+A reactions by taking consistently 
into account all relevant hot and cold nuclear matter effects.
We concentrate on the intermediate and large transverse momentum 
regions ($p_T > 2$~GeV) in minimum bias d+Cu and d+Au and 
central Cu+Cu and Au+Au heavy ion collisions at RHIC with center 
of mass energies of $\sqrt{s} =62.4$~GeV and 200~GeV. We also 
provide the theoretical derivation of the medium-induced 
radiative $\gamma$ spectrum  for a quark emerging out of a 
large $Q^2$ process.

\section{Direct photon production in p+p collisions}
\label{sec:photon@p+p}

\begin{figure*}[t!]
\includegraphics[width=2.in,height=1.in,angle=0]{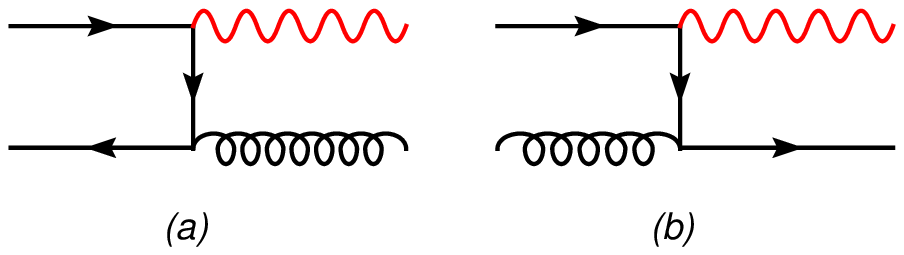} \hspace*{2cm}
\includegraphics[width=2.in,height=1.in,angle=0]{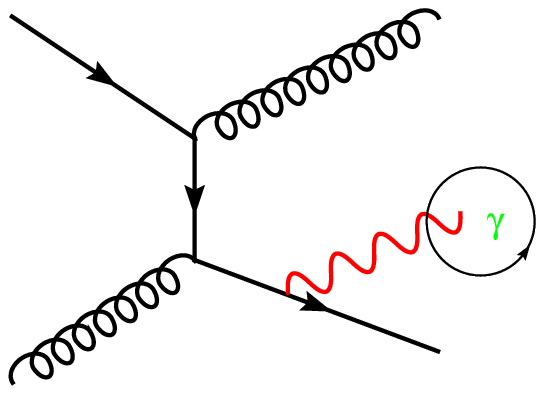}  
\caption[]{Direct photon production from the $q+\bar{q}$
annihilation process + Compton scattering (left panel) and 
bremsstrahlung/fragmentation (right panel).}
\label{fig:prompt}
\end{figure*}

 In ``elementary'' p+p collisions direct photons can be produced, 
at leading-order in perturbation theory, by quark-antiquark annihilation 
 ($q+{\bar q} \rightarrow \gamma g$) and  Compton scattering ($q + g 
\rightarrow \gamma q$), see  LHS of Fig.~\ref{fig:prompt}. One should also 
include the contribution of the bremsstrahlung/fragmentation process 
illustrated in the RHS of Fig.~\ref{fig:prompt}. Naively, it
seems that $\gamma$ radiation gives a higher-order contribution
($\alpha_{em} \alpha_s^2$) when compared to  annihilation and 
Compton scattering ($\alpha_{em} \alpha_s$). However, noticing 
the logarithmic growth with the hard scale $Q^2$ of the photon 
fragmentation function $D_{\gamma/c}(z,Q^2)$, parametrically we have:
\begin{eqnarray}
&&\alpha_s(Q) \propto \ln^{-1}(\frac{Q^2}{\Lambda^2}), \;
D_{\gamma/c}(z,Q^2)\propto \ln(\frac{Q^2}{\Lambda^2}) \nonumber \\
&&\alpha_{em}\alpha_s^2(Q^2) D_{\gamma/c}(z,Q^2) \propto \alpha_{em}\alpha_s(Q^2) .
\end{eqnarray}
Hence the contribution of the bremsstrahlung process is 
effectively the same order as the prompt production processes 
and should be taken into account even in LO calculations.

In our leading-order pQCD model, the cross section of direct photon production
in p+p collision is given by Eq.~(\ref{eq:factorize}), where we also allow for
possible deviations from collinearity for the incoming partons:
\begin{eqnarray}
 &\propto&  \int d^2 {\bf k}_a d^2 {\bf k}_b \; 
{ f({ k}_{a}) f({k}_{b}) } (\cdots) \; .
\label{eq:model-LO}
\end{eqnarray}
Further technical details are given in~\cite{Vitev:2008vk} and we emphasize 
that any overall scale factor (e.g. the process-dependent NLO K-factor) will cancel when
we calculate the nuclear modification ratio $R_{AB}(p_T)$.  

\begin{figure}
\centering
\includegraphics[scale=0.25]{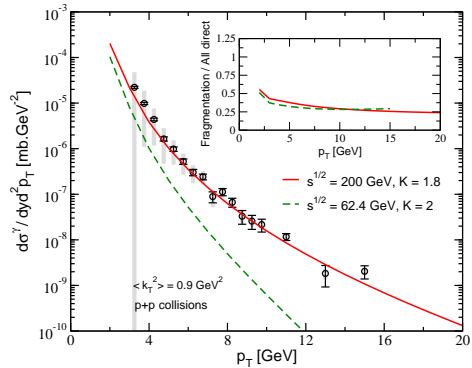}
\caption[]{Cross sections for direct $\gamma$
production in p+p collisions at $\sqrt{s} = 62.4$~GeV  and 200~GeV.
Data at $\sqrt{s}=200$~GeV is taken from PHENIX~\cite{Adler:2006yt}.
Insert gives the ratio of bremsstrahlung photons to all direct
photons.}
\label{fig:photon@pp}       
\end{figure}

In Fig.~\ref{fig:photon@pp} we show our numerical results for 
the differential cross sections for direct photons in p+p collisions at 
$\sqrt{s}=62.4$~GeV and $\sqrt{s}=200$~GeV. The power-law $p_T$
dependence of the data at $200$~GeV, measured by
PHENIX~\cite{Adler:2006yt}, is well described by the pQCD calculation. 
The insert in Fig.~\ref{fig:photon@pp} gives the fraction of 
bremsstrahlung (fragmentation) photons to all
direct photons. We note that at very high $p_T$ the
fragmentation $\gamma$ yield $\sim 25\%-30\%$ of all
direct photons in p+p collisions.

\section{Direct photon production in heavy-ion collisions: 
hot nuclear matter effects}
\label{sec:photon@QGP}

We first discuss final-state QGP-induced modifications to
$d\sigma^{\gamma} / { dy d^2{\bf p}_T  }$. 

\begin{itemize}
\vspace*{0.3cm}
\item{\bf \em Jet quenching in the QGP}
\label{subsec:quenching}

It is well established that when an energetic parton propagates in a
hot/dense nuclear medium, it will suffer multiple scattering and lose
a fraction of its energy via induced gluon radiation~\cite{Gyulassy:2003mc}. 
This jet quenching effect will reduce the contribution of the
fragmentation photons due to the attenuation of the flux of energetic 
quarks. We calculate the parton energy loss using the GLV formalism
and represent the effective fragmentation functions into photons
as follows:
\begin{eqnarray}
 D_{\gamma/c} (z)  \Rightarrow  \int_0^{1-z}
d\epsilon \; P(\epsilon)  \; \frac{1}{1-\epsilon} D_{\gamma/c}
\left( \frac{z}{1-\epsilon} \right) \;\; .
\end{eqnarray}
Here, $P(\epsilon)$ is the probability distribution of the
fractional jet energy loss $\epsilon = \Delta E / E$~\cite{Vitev:2005he}.

\vspace*{0.3cm}
\item{\bf \em Medium-induced photon radiation}
\label{subsec:radiation}

\begin{figure}[!b]
\centering
\includegraphics[scale=0.55]{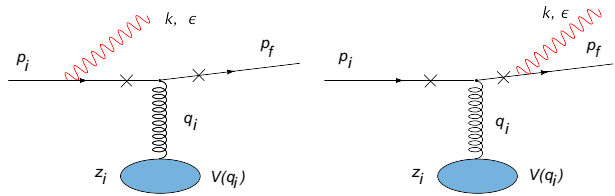}	       
\caption[]{Single-Born diagrams for medium-induced
$\gamma$ emission.}
\label{fig:diags}
\end{figure}

In the hot QGP, a propagating jet may also radiate  photons
due to multiple scattering in the medium~\cite{Vitev:2008vk,Srivastava:2008es}
Because there is no gauge-boson self-interactions in $\gamma$ bremsstrahlung, 
one may expect that the $dI^\gamma/d\omega$ evaluation is a trivial 
simplification of  the gluon intensity calculation. 
This naive expectation is not true. Consider the radiative
amplitude for single scattering of a fast on-shell
quark (there are 3 diagrams in QCD, while 2 diagrams in QED as shown 
in Fig.~\ref{fig:diags}):
\begin{equation} {\cal M}_{rad}^{\gamma}(k) \propto  2  i g_s {\bf \epsilon}_\perp
\cdot \bigg( \frac{ {\bf k}_\perp}{{\bf k}_\perp^2}  - \frac{ ({\bf
k}-{\bf q})_\perp}{({\bf k}-{\bf q})_\perp^2 } \bigg) e^{i\frac{{\bf
k}_\perp^2}{2k^+} z^+ } [T^c,T^a] \; .  \; 
\end{equation} 
In the QED limit $T^{c}\rightarrow 1$ and we obtain ${\cal
M}_{rad}^\gamma(k) \rightarrow 0 $. Thus, a theoretical
approach optimized for medium-induced gluon radiation cannot
be directly applied to photon bremsstrahlung in the medium and one
should be careful in identifying the correct kinematic approximations 
in evaluating $\gamma$ emission~\cite{Vitev:2008vk,Zhang:2003}.

Meaningful direct photon phenomenology requires theoretical advances
in understanding the medium-induced $\gamma$ radiation off of quarks produced 
inside the plasma in large $Q^2$ processes. We derive this contribution
to the $\gamma $ spectrum using
the Reaction Operator  approach~\cite{Vitev:2008vk,Gyulassy:2003mc}. 
For the single-Born scattering diagrams, illustrated in Fig.~\ref{fig:diags}, 
we obtain:
\cite{Vitev:2008vk}
\begin{equation}
{\cal M}_{rad}(k,\{i\}) = e\left( \frac{\epsilon \cdot p_f}{k \cdot
p_{f}} - \frac{\epsilon \cdot p_{i}}{k \cdot p_{i}}  \right) e^{i
z_i^+ k^-} \; , \label{hard}
\end{equation}
where the collisional part of the amplitude is not shown. The contribution of
Double-Born diagram is found to be negligible: $ {\cal M}^V_{rad}(k)  
\approx 0 \; . $  
Extending the above calculations to higher orders in the correlation 
between the multiple scattering centers we 
demonstrate that contributions to $dI^{\gamma}/d\omega$ vanish 
beyond second order in
opacity~\cite{Vitev:2008vk}:
\begin{eqnarray} &&k^+ \frac{dN^\gamma (k)}{dk^+ d^2{\bf k}_\perp} =
\frac{\alpha_{em}}{\pi^2} \bigg\{ \int\frac{d \Delta
z_1}{\lambda_q(z_1)} \int d^2 {\bf q}_{\perp \, 1}
\frac{1}{\sigma^{\rm el}}\frac{d^2 \sigma^{\rm el}}{ d^2 {\bf
q}_{\perp \,1 } }
\nonumber \\
&& \times \left[ |{\cal M}_{rad}(\{1\})|^2 +2 {\cal
M}^*_{rad}(\{1\}) {\cal M}_{rad}(\{0\}) \cos( k^- \Delta z_1^+)
\right] \nonumber \\ & & + \,\,  \texttt{corrections} \,.
\label{eq:photon@final}
\end{eqnarray}
Here $\Delta z_i^+ = z_i^+ - z_{i-1}^+$, and $\lambda_q$ is the quark mean free path.
We define the inverse photon
formation time as $ \tau_f^{-1} =  {\bf
k}^2/(2 \omega) \approx \surd{2}\, k^-$. There are two well-defined  limits in Eq.~(\ref{eq:photon@final}):
when $\tau_f^{-1} \lambda_q \gg 1$, the term   $\propto \cos( k^-
\Delta z_i^+)$  vanishes due to oscillation and we  recover incoherent $\gamma$ 
emission. In the opposite case,  $\tau_f^{-1} \lambda_q \leq 1$, interference
between the vacuum and medium-induced photons becomes critical.

\vspace*{0.3cm}
\begin{figure}[ht]
\centering
\includegraphics[scale=0.25]{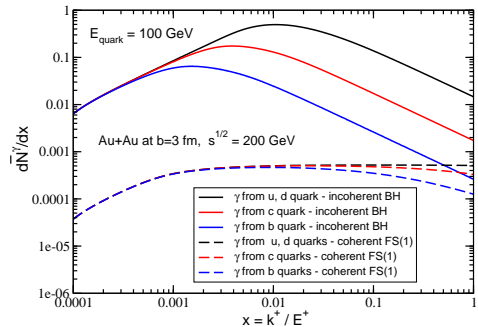}	       
\caption[]{Scaled medium-induced photon number
spectrum versus $x=k^+/E^+$ for $E_q = 100$~GeV light, charm and
bottom quarks in central Au+Au collisions at $\sqrt{s} = 200$~GeV.
Solid curves represent the incoherent Bethe-Heitler (BH) 
photon radiation in the medium, and dashed curves stand for 
the coherent final-state (FS) photon 
bremsstrahlung.}
\label{fig:photon@radiation}
\end{figure}
In Fig.~\ref{fig:photon@radiation} we show  numerical results for
the scaled medium-induced photon number spectrum $\bar{ d {N}^\gamma}/dx
= (e/e_q)^{2}  d{N}^\gamma/dx $, $x = k^+ / E^+$ for a quark jet
propagating outwards from the center of the medium created in 
Au+Au collisions at RHIC with the impact
parameter $b=3$~fm . It is apparent that
interference effect strongly suppresses  stimulated photon
emission when compared to the incoherent scenario.
This reduction in emission strength, which is relevant for photon production 
in heavy ion reactions, differs significantly from the argued 
small LPM effect for on-shell  quarks, see e.g.~\cite{Srivastava:2008es,Turbide:2005}.  
In our detailed numerical simulations~\cite{Vitev:2008vk,Vitev:2005he} we also find 
that the average number of jet interactions in the medium is never large, 
$\langle n \rangle = L / \lambda_q = 2-3 $.

\vspace*{0.3cm}
\item{\bf \em Jet-photon conversion}
\label{subsec:conversion}

In the hot/dense medium, a possible new source of direct $\gamma$
is jet-photon conversion~\cite{Fries:2003,Srivastava:2008es}, see e.g. 
Fig.~\ref{fig:jet-conversion}. Making the approximation that $p_\gamma \approx
p_c$ in the forward scattering process, we can derive the additional differential 
photon multiplicity  as follows: 
\begin{equation}
 N^\gamma_{\rm conv.}(c) =
\int_{t_0}^L dt \; \rho(T) \sigma^{qg\rightarrow \gamma q}_{tot} (T)
\;, \label{eq:conversion}
\end{equation} where the cross section is given by
\begin{equation}
\sigma^{qg\rightarrow \gamma q } = \frac{\pi \alpha_s \alpha_{em}}{6
m_D E} \ln\frac{E}{2m_D} \; , \end{equation}
with $s \approx 2 m_D E$ and $t \in (m_D^2, s/4)$. Here $m_D =g_s T$ is
the Debye screen mass.

\begin{figure}[!b]
\centering
\center{\resizebox{0.38\textwidth}{!}{%
 \includegraphics{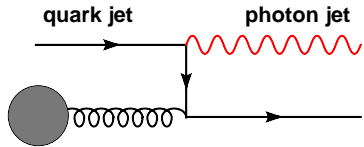}
} }\vspace*{0.2cm} \caption{A fast quark jet is converted into a high
energy photon jet via an interaction with a thermal gluon in the hot QGP. }
\label{fig:jet-conversion}
\end{figure}

\vspace*{0.3cm}
\item{\bf \em Total QGP contribution to direct photons}

Combining our results, we obtain the net effect on direct $\gamma$ 
productions from the final-state jet-medium interactions in the QGP:
\begin{eqnarray}
D_{\gamma/c} (z) & \Rightarrow & \int_0^{1-z} d\epsilon \;
P(\epsilon)  \; \frac{1}{1-\epsilon} D_{\gamma/c} \left(
\frac{z}{1-\epsilon} \right) \nonumber \\  && +  \,
\frac{dN^\gamma_{\rm med.}(c)}{d z}  \, + \, N^\gamma_{\rm conv.}(c)
\delta(1-z) \;, \quad \label{eq:photon@hot}
\end{eqnarray}
where the first term takes into account the jet quenching effect on
fragmentation photons, the second term $ \frac{dN^\gamma_{\rm med.}(c)}{d z}$
stands for the
additional contribution from medium-induced photons which
can be derived from Eq.~(\ref{eq:photon@final}), and 
the last term gives the correction from the inverse Compton 
scattering of the fast quark in the QGP.

\end{itemize}

\section{Direct photon production in heavy-ion collisions: 
cold nuclear matter effects}
\label{sec:photon@cold}

Even though in p+A collisions there are no final-state hot nuclear
medium effects, these still can not be regarded as a
simple superposition of p+p scatterings. There are different types of  
CNM interactions that will manifest themselves in the cross section for
direct photon production in p+A  and A+A  reactions.

\begin{itemize}

\item{\bf \em Isospin effect}

The cross sections for direct photon production for p+p, p+n and n+n
collisions are different because they depend on the electric charges of the quarks 
($\sigma \propto \sum_q e_q^2 $). The different quark composition of $p$ and 
$n$ will have a significant impact on the nuclear modification factor of 
direct photon production $R_{AB}^\gamma(p_T)$ if the large $x_B$ region in the PDFs
is probed.

\item{\bf \em Initial-state energy loss in cold nuclei}

A fast parton passing through cold
nuclear matter may also lose energy before the hard scattering.  There is 
plentiful evidence that this  initial-state energy loss is  important for heavy
ion phenomenology but only recently has it been calculated~\cite{Vitev:2007ve}.
Its effect can be modeled  as:
\begin{eqnarray}
\phi_{a,b/N}({x}_{a,b}, Q^2) &\rightarrow & \phi_{a,b/N}
\left( \frac{ x_{a,b} }{1-\epsilon_{a,b}}, Q^2 \right) 
 \;  , \quad\label{eq:EL@cold}
\end{eqnarray}
where $\epsilon_{a,b}$ are the fractional energy losses for the
incoming partons  $a, b$ evaluated in the rest frame of the
corresponding target nucleus with, typically, $\epsilon_{a,b} \ll 1$.

\item{\bf \em Cronin effect}

Initial-state multiple scattering in cold nuclear matter will
broaden the transverse momenta of incoming partons before the hard
scattering. The Cronin effect, in our calculation, is modelled via such  
$k_T$ diffusion, $\langle k_T^2 \rangle = \langle k_T^2 \rangle_{pp}
 + \langle k_T^2 \rangle_{med}$ ~\cite{Gyulassy:2003mc},
\begin{eqnarray}
&& \langle k_T^2 \rangle_{med} =
 \left(\frac{2\mu^2 L}{\lambda}\right)_{q,g} \zeta  \;, 
\end{eqnarray}
where $\mu^2 = 0.12$~GeV$^2$, $\lambda_g = (C_F/C_A) \lambda_q =
1$~fm, $L$ denotes the length of the nuclear medium 
and $\zeta$ accounts for the logarithmic dependence arising 
from large-angle scattering.  Eq.~(\ref{eq:model-LO}) allows one to 
incorporate easily such transverse momentum broadening into the
 pQCD calculation.

\item{\bf \em Shadowing and EMC effect}

In our model shadowing  was calculated from the coherent
final-state parton interactions within the pQCD higher-twist collinear
factorization approach~\cite{Vitev:2008vk}. The scale of
power correction   $\xi^2$ is the only parameter constrained by
the mean squared momentum transfer per unit length $\mu^2/\lambda$,
$ (\xi^2 A^{1/3})_{q,g} \approx ( 2 \mu^2 L/ \lambda )_{q,g}$ in
minimum bias collisions, and by the world's data on DIS on nuclei.
Only the EMC effect is parametrized in our numerical simulations.

\end{itemize}

\section{Numerical results}
\label{sec:numerical}

\begin{figure}
\vspace*{0.3cm}
\centering
\resizebox{0.48\textwidth}{!}{%
  \includegraphics{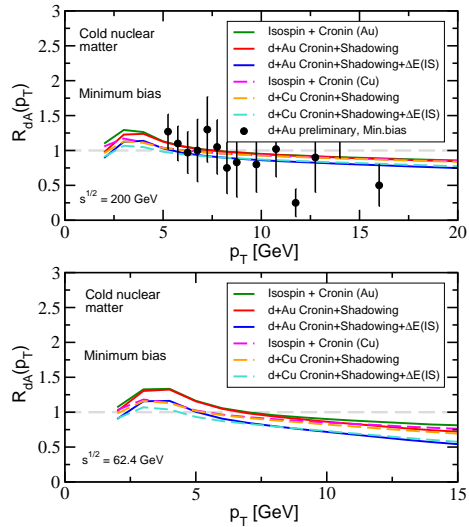}
}
\vspace*{0.3cm}
\caption[]{ Nuclear modification factors $R_{dAu}^{\gamma}(p_T)$ for direct
$\gamma$ production in minimum bias d+Au (solid lines) and d+Cu (dashed
lines) collisions at $\sqrt{s} = 62.4$~GeV (bottom panel) and
200~GeV (top panel). Preliminary $\sqrt{s} = 200$~GeV minimum
bias d+Au data is from PHENIX~\cite{Peressounko:2006qs}.}
\label{fig:photon@DA}       
\end{figure}

With the leading-order pQCD improved parton model in
Eqs.~(\ref{eq:factorize}) and (\ref{eq:model-LO}),  augmented 
by including all QGP effects,
calculated in Sect.~\ref{sec:photon@QGP}, and CNM
effects, discussed in Sect.~\ref{sec:photon@cold}, we can evaluate
numerically the nuclear modification factor
$$ R_{AB}^{\gamma}(p_T,b) = \frac{d\sigma_{AB}}{dyd^2{\bf p}_T} \bigg/
N_{AB}^{\rm coll}(b)\frac{d\sigma_{pp}}{ dyd^2{\bf p}_T } \; , $$
and compare to available experimental data at RHIC. Since we
focus on hard photon production with $p_T > 2-3$~GeV,  
the contribution from thermal photons 
is very small 
and thus neglected in the current study.

Figure~\ref{fig:photon@DA} shows  numerical results 
of direct photon production for
minimum-bias d+Au and d+Cu collisions at different colliding
energies $\sqrt{s} = 62.4,\;  200$~GeV by taking into 
account all relevant CNM effects, which has not
yet been done before in the literature according to our
knowledge. Here to demonstrate the
relative importance of different CNM effects, we
calculated $R_{dA}^{\gamma}(p_T)$
under different theoretical assumptions, namely, 
$R_{dA}^{\gamma}(p_T)$ with isospin and Cronin effects, 
$R_{dA}^{\gamma}(p_T)$
with isospin, Cronin effect and shadowing effects, and 
$R_{dA}^{\gamma}(p_T)$ with the initial-state energy loss
as well as isospin, Cronin and shadowing effects. And in all 
numerical simulations the EMC effect is already included.
We note that when $p_T <
6$~GeV, the Cronin effect is dominant and leads to cross section  
enhancement. For $p_T > 6$~GeV, the isospin effect becomes important
and $R_{dA}^{\gamma}(p_T)< 1 $. The effect of initial-state energy loss 
on direct $\gamma$ attenuation is substantial but 
the EMC effect is small and noticeable only at
the low colliding energy and at the largest transverse momenta.
 It is interesting to note that CNM effects  have reduced
direct photon production at $p_T \sim 15$~GeV by about $25\%$ for
d+A collisions at $\sqrt{s}=200$~GeV, and by about $40\%$ for reactions
at $\sqrt{s}=62.4$~GeV. Due to large error bars in the data, current 
experimental measurements in d+A cannot  constrain more 
quantitatively parton dynamics in large nuclei and progress in
this area is urgently needed.

\begin{figure}
\vspace*{0.3cm}
\centering
\resizebox{0.48\textwidth}{!}{%
  \includegraphics{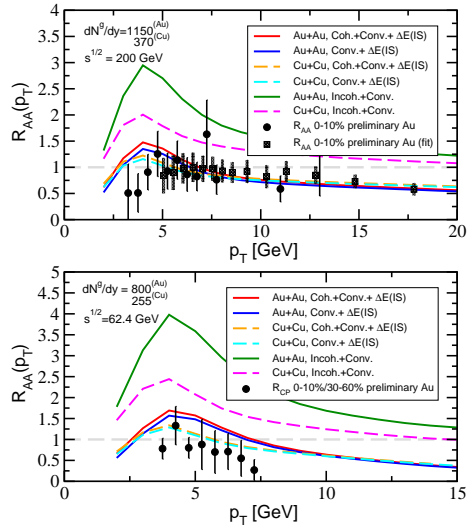}
} 
\caption[]{$R_{AA}^{\gamma}(p_T)$ for direct photon in central Au+Au and
Cu+Cu collisions at $\sqrt{s}=62.4$~GeV (bottom panel) and
$\sqrt{s}=200$~GeV (top panel) calculated including cold and hot nuclear  matter 
effects. Data are taken from Refs.~\cite{takao,Adler:2006yt}.
}
\label{fig:photon@AA2}       
\end{figure}

Next, we show  results of direct photon production 
in central Au+Au and Cu+Cu collisions at
$\sqrt{s}=62.4$~GeV and $200$~GeV in  Fig.~\ref{fig:photon@AA2}.
Similar to calculations for $R_{dA}^{\gamma}(p_T)$, in Fig.~\ref{fig:photon@AA2}
we demonstrate the relative strength of different nuclear matter effects
by giving three curves of $R_{AA}^{\gamma}(p_T)$: one stands
for $R_{AA}^{\gamma}(p_T)$ with coherent medium-induced 
photon bremsstrahlung and jet-photon conversion as well as initial-state
energy loss; another curve represents $R_{AA}^{\gamma}(p_T)$ with
jet-photon conversion and initial-state energy loss; and the third curve 
denotes $R_{AA}^{\gamma}(p_T)$ with only incoherent
photon radiation and jet-photon conversion in the QGP.
By comparing Fig.~\ref{fig:photon@AA2} to Fig.~\ref{fig:photon@DA} 
one can see that in A+A collisions $R_{AA}^{\gamma}(p_T)$ for direct photon production 
is dominated by cold nuclear matter effects, amplified by the presence of 
two large nuclei. We note that in our treatment of photon bremsstrahlung in
the QGP we have considered the coherent interference of medium-induced photon
radiation with the hard emission from the large $Q^2$ scattering of the parent
quark, which effect, as well as the initial-state energy loss effect, 
has been ignored in a previous study~\cite{Turbide:2005}.   
Experimental data clearly exclude both large Cronin 
enhancement and incoherent photon emission,  thereby providing support for
the new theoretical developments reported in~\cite{Vitev:2008vk,Vitev:2007ve}.   
Copious jet-photon conversion~\cite{Fries:2003}  is also ruled out. 
In fact, in contrast to early 
speculations we find that  $p_T < 5$~GeV medium-induced photons contribute  
$\sim 10\%$ to the observed cross section  and inverse Compton scattering 
of energetic quarks in the QGP  is a $< 25\%$ correction.
In the high $p_T$ region, the two enhancement contributions are very small and
$R_{AA}^{\gamma}(p_T)< 1$.


\section{Conclusions}
\label{sec:conclusions} 

By taking into account all relevant cold nuclear matter effects
and the hot QGP effects we carried out the first systematic 
study of direct photon production in p+A and A+A collisions at 
RHIC~\cite{Vitev:2008vk}. After consistent modeling of initial- and
final-state  parton propagation in the medium we did not find 
theoretical justification for the previously argued large QGP-induced 
enhancement of the direct photon cross section. In fact, our 
results suggest that in both proton-nucleus and nucleus-nucleus 
reactions cold nuclear matter effects are dominant. Experimental
data is compatible with such an interpretation 
($R_{AB}^{\gamma}(p_T) \sim 1$) and provides support for the theoretical 
underpinnings~\cite{Gyulassy:2003mc,Vitev:2008vk,Vitev:2007ve} of our 
numerical results.

\section*{Acknowledgments}

This research is supported by the US Department
of Energy, Office of Science, under Contract No. DE-AC52-06NA25396
and in part by the LDRD program at LANL, the
MOE of China under Project No. IRT0624 and the NNSF of China.



\begin{thebibliography}{0}
\bibitem{Gyulassy:2003mc}
  M.~Gyulassy, I.~Vitev, X.~N.~Wang and B.~W.~Zhang,
  arXiv:nucl-th/0302077.


\bibitem{VWZ}
I.~Vitev, S.~Wicks and B.~W.~Zhang,
  JHEP {\bf 0811} (2008) 093
  [arXiv:0810.2807 [hep-ph]].


\bibitem{Srivastava:2008es}
D.~K.~Srivastava,
  J.\ Phys.\ G {\bf 35} (2008) 104026
  [arXiv:0805.3401 [nucl-th]].



\bibitem{takao}
T.~Sakaguchi,
  J.\ Phys.\ G {\bf 35} (2008) 104025
  [arXiv:0805.4644 [nucl-ex]].

\bibitem{Vitev:2008vk}
  I.~Vitev and B.~W.~Zhang,
  Phys.\ Lett.\  B {\bf 669} (2008) 337
  [arXiv:0804.3805 [hep-ph]].


\bibitem{Fries:2003}
R. J. Fries, B. M\"uller and D. K. Srivastava, Phys.
Rev. Lett. {\bf 90}, 132301 (2003).

\bibitem{Adler:2006yt}
  S.~S.~Adler {\it et al.}  [PHENIX Collaboration],
  Phys.\ Rev.\ Lett.\  {\bf 98}, (2007) 012002.

\bibitem{Vitev:2005he}
  I.~Vitev,
  Phys.\ Lett.\  B {\bf 639}, 38 (2006).


\bibitem{Zhang:2003}
  B.~W.~Zhang and E.~K.~Wang,
  Chin.\ Phys.\ Lett.\  {\bf 20} (2003) 639.

\bibitem{Turbide:2005}
S. Turbide, C. Gale, S. Jeon and G. D. Moore, Phys. Rev.
C {\bf 72}, 014906 (2005).

\bibitem{Vitev:2007ve}
  I.~Vitev,
  Phys.\ Rev.\  C {\bf 75}, 064906 (2007).

\bibitem{Peressounko:2006qs}
  D.~Peressounko  [PHENIX Collaboration],
  Nucl.\ Phys.\  A {\bf 783}, 577 (2007).


\end{thebibliography}
\end{document}